\documentclass[aps,prl,nofootinbib,floatfix,twocolumn,letterpaper]{revtex4}

\usepackage{graphicx, color, upgreek}
\usepackage{bm}
\usepackage{amsmath,amssymb,amsfonts}

\newcommand{\HH}{\mathcal{H}}

\begin{document}

\title{Testing general relativity on accelerators}

\author{Tigran Kalaydzhyan}
\email[]{tigran.kalaydzhyan@desy.de}
\affiliation{Department of Physics and Astronomy, Stony Brook University,\\ Stony Brook, New York 11794-3800, USA}

\begin{abstract}
Within the general theory of relativity, the curvature of spacetime is related to the energy and momentum of the present matter and radiation.
One of the more specific predictions of general relativity is the deflection of light and particle trajectories in the gravitational field of
massive objects. Bending angles for electromagnetic waves and light in particular were measured with a high precision. However, the effect of
gravity on relativistic massive particles was never studied experimentally. Here we propose and analyze experiments devoted to that purpose.
We demonstrate a high sensitivity of the laser Compton scattering at high energy accelerators to the effects of gravity.
The main observable -- maximal energy of the scattered photons -- would experience a significant shift in the ambient gravitational
field even for otherwise negligible violation of the equivalence principle. We confirm predictions of general relativity for
ultrarelativistic electrons of energy of tens of GeV at a current level of resolution and expect our work to be a starting point
of further high-precision studies on current and future accelerators, such as PETRA, European XFEL and ILC.
\end{abstract}


\maketitle

{\it Introduction.---}
Einstein's theory of general relativity (GR) is a well established theory of gravity confirmed in all observations and experiments to date~\cite{Will:2014xja}.
One of the classical and essential tests of GR is based on the gravitational light bending in the presence of a large massive body. Measurements of the light bending deflection started from a spectacular observation of starlight deflection during a solar eclipse about a century ago~\cite{eddington}, and were expanded to
radio-waves becoming very precise~\cite{Lebach:1995zz, Shapiro:2004zz}.
For the Earth, however, a direct experimental measurement of the gravitational bending remains infeasible because of the smallness of the expected deflection \cite{Landau1975}, about 3~nrad:
\begin{equation}
\theta_\oplus=\frac{4 G M_\oplus}{c^2 R_\oplus} \approx 2.78\times 10^{-9},
\label{Eangle}
\end{equation}
where $G$ is the gravitational constant and $c$ is the speed of light (which we will put to $c=1$ in order to work in natural units).
The bending magnitude for the light generated and studied in a ground based experiment will be even
smaller,
\begin{equation}
\theta_{\mathrm{lab}}=\frac{2 G M_\oplus}{c^2 R_\oplus}\left(\frac{L}{R_\oplus}\right),\qquad L \ll R_\oplus,  
\label{Lab-angle}
\end{equation}
where $L$ is the length of the light trajectory.
Thus, for a distance of 1~m, this angle is only $2\times 10^{-16}$~rad and
the light shifts by 0.2~fm, which is undetectably small, at least for a
direct measurement. However, the bending property of gravity can be used to test its effect on the
relativistic massive particles (which is not known neither from the ground based experiments, nor from astrophysical observations) through an effective refractive index. The equivalence principle predicts impossibility of any of such Earth-based tests, since all processes will be the same as in the absence of the Earth's (or any other ambient) gravitational field. However, there are no proofs of the equivalence principle at high energies either.

In this article, we describe a laboratory method that tests the validity of the equivalence principle at high energies utilizing the concept of the gravitational bending.
In brief, we consider the high-energy Compton scattering, where the main observable -- maximal energy of the scattered photons
(Compton edge) -- is extremely sensitive to the (relative) gravitational bending angle of the photon and an electron, even in the Earth's weak
gravitational field. In some sense, this is an ultrarelativistic version of the famous E\"otv\"os experiment \cite{eotvos, Wagner:2012ui},
where even a small difference in the action of gravity on different materials would cause a noticeable change in the torsion pendulum orientation.
In our case, even a tiny deviation between the action of gravity on the photon and electron (besides the one prescribed by GR) would result
in a significant shift of the Compton edge.

{\it Gravity as a bending medium.---}
In order to study the effect of gravity on the relativistic particles we use an elegant
reformulation of Riemannian geometry in terms of optics of continuous medium in a flat space.
The idea is that the gravity bends trajectories of particles in a way that mimics the presence
 of a nontrivial effective refractive index. This idea going back to the early ages of general relativity \cite{Eddington},
 was suggested by Einstein himself and was employed by many authors, see Ref.~\cite{de Felice:1971ui}
and Refs. therein. Let us demonstrate how such refractive indices for the photon, $n$, and electron,
$n_e$, can be derived. The Earth's gravitational field, i.e. static weak field of a spherically symmetric body, can be described by the isotropic metric,
\begin{align}
ds^2 = \HH^2 dt^2 - \HH^{-2}(dx^2 + dy^2 + dz^2),\label{metric} 
\end{align}
where, on the surface of the Earth,
\begin{align}
\HH \equiv \sqrt{1-\frac{2 G M_\oplus}{R_\oplus}}\,.
\end{align}
Unless noted otherwise, we do not consider gravitational potentials from other celestial bodies. We will return to this discussion at the end of the paper.
Trajectory of a photon (light) is described by null geodesics, $ds^2=0$, which leads to the expression on the coordinate speed of light $v_\gamma$ and the refractive index $n$,
\begin{align}
\frac{c}{n}=v_\gamma  \equiv \biggl | \frac{d\vec x}{dt}\biggr | = \HH^2. 
\end{align}
In other words, this introduces an effective dispersion relation for the photon, $k = n \omega$ with
\begin{align}
n=1+\frac{2 G M_\oplus}{ R_\oplus} \approx  1+1.39\times 10^{-9}\,.\label{refearth}
\end{align}
The latter expression has also been derived by other
authors, see, e.g.,~\cite{grav=optics, Boonserm:2004wp, Ye:2007df}. For a massive probe particle moving with the coordinate speed $v_m$, the line element can be rewritten then as
\begin{align}
ds^2 = \HH^2 \left( 1- n^2 v_m^2 \right) dt^2,
\end{align}
and the relativistic action takes the form
\begin{align}
S = -\int m\,ds = -\int m \HH \sqrt{1-n^2 v_m^2} dt\,.
\end{align}
Using this action, one can easily obtain the coordinate momentum $\vec p$ and the Hamiltonian (energy) ${\cal E}$,
\begin{align}
\vec p = \frac{m \HH n^2}{\sqrt{1-n^2 v_m^2}}\, \vec v_m, \qquad {\cal E} = \frac{m \HH}{\sqrt{1-n^2 v_m^2}}\,.
\end{align}
The electron refractive index, $n_e \equiv p / {\cal E}$, is then
\begin{align}
n_e = n^2 v_m = n \sqrt{1 - \frac{m^2 \HH^2}{{\cal E}^2}}\,. \label{ne}
\end{align}
One has to keep in mind that the physical observables in GR for the metric (\ref{metric}) should be obtained by rescaling of the coordinate ones (e.g., $\tilde v = \HH^{-2} v$, $\tilde p = \HH p$, $\tilde {\cal E} = \HH^{-1} {\cal E}$). 
In this case, the physical dispersion relations for the photon, $\tilde k = \tilde \omega$, and for the electron, $\tilde {\cal E}^2 = \tilde p^2 + m^2$, have the same form as in the absence of gravity. However, if the Newton's constant for the relativistic electron, $G_e$, and the photon, $G_\gamma$, are different, at least one of the dispersion relations should be modified. For the sake of simplicity, we perform calculations with the coordinate observables. One can substitute the physical values for the observables in the final expressions, since ${\HH \approx 1}$ within considered precision.

The main difference between the gravitational ``medium" presented
by Eq.~(\ref{refearth}) and any material medium
is the bending independence on frequency of light or photon energy,
which is a consequence of the gravity geometrical interpretation or
the curved space-time concept.
Another important difference is that the optical refractive index is a property of light only, while the gravity-induced index is a property of any kind of matter or radiation. This manifests itself in, e.g., absence of the vacuum Cherenkov radiation induced by gravity, as one can easily deduce from Eq.~(\ref{ne}), i.e. there is never a situation when $v_m > v_\gamma$.

{\it The Compton process in a gravitational field.---}
High energy Compton scattering describes a photon of energy $\omega_0$ which scatters off an electron with mass $m$ and energy ${\cal E}\gg m$ and acquires an energy $\omega \gg \omega_0$.
In order to demonstrate the sensitivity of the process to gravity, let us assume for a moment that $n_e=\beta\equiv\sqrt{1-m^2/{\cal E}^2}$, i.e. electrons are not attracted to the Earth while being attracted by other celestial bodies.
Then, from
the energy-momentum conservation,
with $n\approx 1$, we derive
\begin{align}
n-1 = \frac{m^2}{2 {\cal E} ({\cal E}-\omega)}\left( 1+x+\left(\frac{{\cal E}-\omega}{m} \right)^2\theta^2 - x \frac{{\cal E}}{\omega}\right),
\end{align}
where $x=4{\cal E} \omega_0\sin^2{(\theta_0/2)}/m^2$,
with the initial photon's angle being denoted
by $\theta_0$, while the scattered photon angle is $\theta \ll 1$; the angles are defined relative to the electron.
This kinematic expression is derived for small refractivity and high energies,
i.e., the  $\mathcal{O}((n-1)^2)$, $\mathcal{O}(\theta^3)$, and
$\mathcal{O}((m/{\cal E})^{3})$ terms are neglected.
The formula allows us to find the maximal energy $\omega=\omega_{max}$
of the scattered photons (so-called Compton edge, at $\theta=0$) in the Earth's gravitational
field.
\begin{figure}[t]
\centering
\includegraphics[width=8cm]{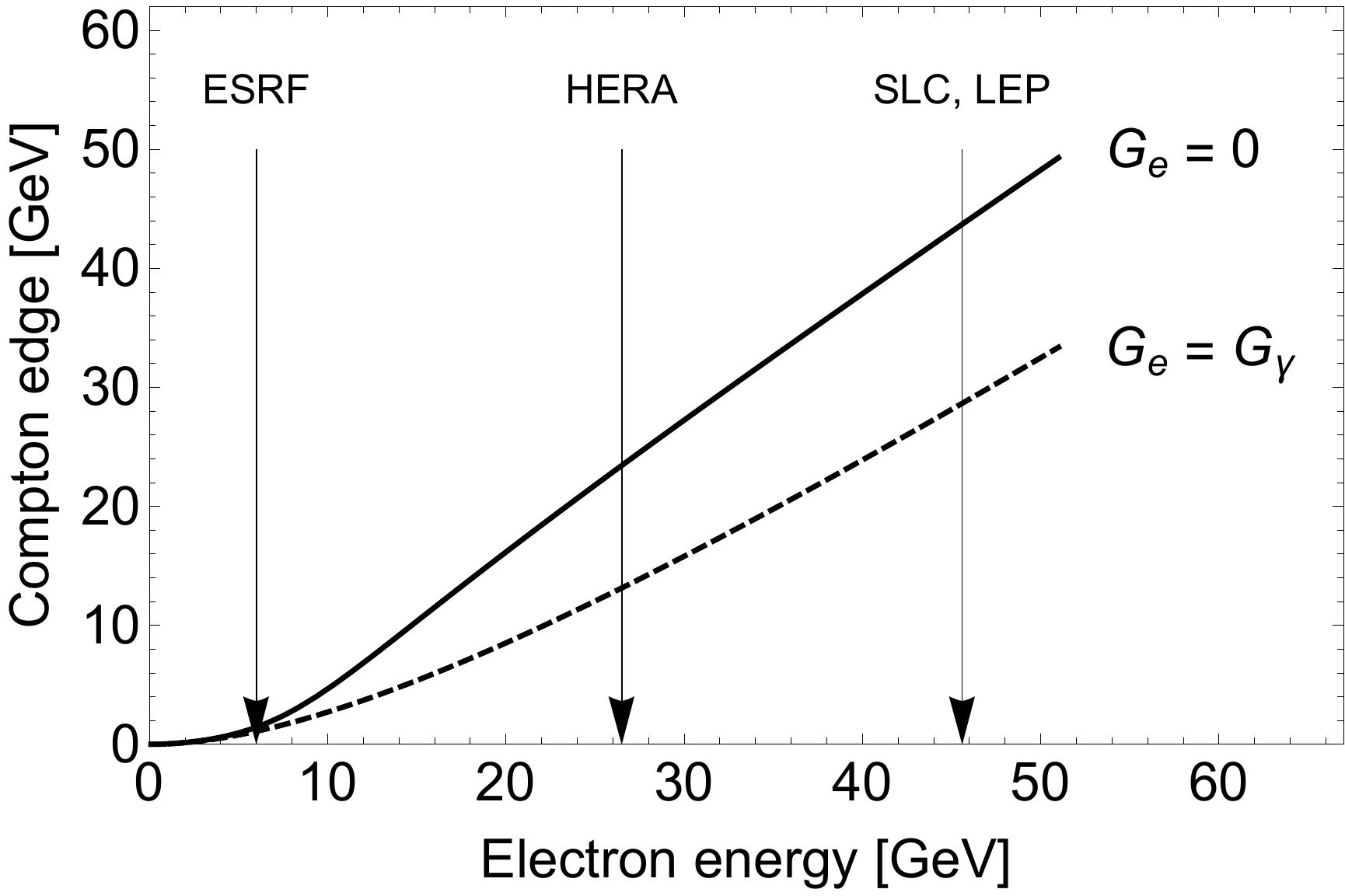}
\caption{\label{fig1}
Compton scattered photons' maximal energy (Compton edge)
dependence on the initial electron energy
for a head-on collision with 514.5~nm laser light.
Solid and dashed lines correspond to the situations $G_e = 0$ (i.e. $n_e = \beta$) and $G_e = G_\gamma$ , respectively.}
\end{figure}

Now, if the gravity affects relativistic electrons as well, in the way prescribed by Eq.~(\ref{ne}), then the electron momentum is $p=n_e\mathcal{E}$, and the same derivation from the energy-momentum conservation will give us the Compton edge,
\begin{align}
\omega_{max} = \frac{\mathcal{E}x}{1+x}\,,\label{nominal}
\end{align}
for $any$ refractive index $n$ close to unity, which is a natural consequence of the equivalence principle, i.e., within GR, it is impossible to observe the effect of gravity in a free-falling system.

To compare the resulting dependencies for an intentionally wrong case, $G_e=0$, and the case predicted by GR (\ref{ne}), we take $\omega_0=2.41~\mathrm{eV}$ (green, 514~nm, Argon ion laser), $\theta_0 \approx \pi$,
and various energies of the accelerator electrons, see Fig.~\ref{fig1}.
The plot shows considerable sensitivity, which grows toward high energies
in a range available to accelerating laboratories.

The complete absence of the effect of gravity on the electron is, of course, unrealistic and was used for demonstration purposes only. We expect deviations from GR to be subtle, if present, since otherwise they would have been observed in the Compton scattering experiments long ago. In order to quantify the possible deviations, we can introduce a correction $\delta_e = \delta_e({\cal E}, m, \cal{H})$,
\begin{align}
n_e = n \sqrt{1 - \frac{m^2}{{\cal E}^2}}+ \delta_e \label{newne}\,,
\end{align}
with $10^{-18}\ll\delta_e< 10^{-9}$, where the left bound indicates that we do not have to consider subleading (Newtonian) term
 in ${\HH}$ in (\ref{ne}) and we can perform calculations without going to the next orders of small parameters\footnote{One can show that the case of $\delta_e\neq 0$ is equivalent to the isotropic Lorentz-violation in the QED sector of the Standard Model Extension~\cite{sme} via $c_{00}=3c_{jj}=3\delta_e /4$ (no summation by $j$).}. We also assume
that $\delta_e$ does not change by many orders of magnitude with small changes of its arguments and satisfies existing low-energy experimental
limits. We do not imply any functional form of $\delta_e$ and focus only on possible experimental detection of $\delta_e\neq 0$.
Physically, e.g., $\delta_e > 0$ would mean that the electron is coupled to the gravity stronger than predicted by GR.
Quantum corrections from the $G\gamma\gamma$ vertex \cite{Berends:1975ah, Delbourgo:1973xe} would contribute to $n$ but not to $\delta_e$ and will cancel out, while corrections to $G e^+ e^-$ vertex will be negligibly small \cite{Berends:1975ah} to contribute to $\delta_e$ at given precision, so we ignore (expected) quantum gravity effects.
Assuming a shift $\Delta\omega \ll \omega_{max}$ in the Compton edge,
\begin{align}
\omega_{max}= \frac{\mathcal{E}x}{1+x} - \Delta\omega\,,
\end{align}
and using energy-momentum conservation condition, we obtain a relation between $\delta_e$ and $\Delta\omega$,
\begin{align}
\delta_e = \frac{m^2(1+x)^3}{2{\cal E}^3 x}\Delta\omega\,. \label{deviation}
\end{align}
If one absorbs $\delta_e$ in the definition of Newton's constant for the electron, then the difference $\Delta G = G_e-G_\gamma \ll G$ is
related to the shift of the Compton edge by
\begin{align}
 \frac{\Delta\omega}{\omega_{max}} = 2\gamma^2 \cdot\frac{n - 1}{(1+x)^2}\cdot \frac{\Delta G}{G}\,,\label{wtog}
\end{align}
which makes it more evident that it is the electron's large Lorentz factor $\gamma$ which reveals possible deviations from GR in the experiment. Alternatively, one can keep $G$ as a universal coupling and absorb $\delta_e$ into the definition of the gravitational mass of electron, $m_g = m_e + \Delta m$. In this case $\Delta G/G$ should be replaced by $\Delta m/m_e$ in the formula above, as well as in Fig.~\ref{figne}.

{\it Experimental results.---}
The high-energy accelerators with laser Compton facilities, such as ESRF, HERA, SLC and LEP have been used in particle physics studies for years (see their energy and $x$-parameters in Table~\ref{tab1}).
\begin{table}[b]
\begin{tabular}{|c|c|c|c|c|}
\hline
            & Electron & Kinematic & $\omega_{max}$    & $\omega_{max}$ \\
Accelerator & energy & factor      &    [$G_e=G_\gamma$]     & [$G_e = 0$] \\
            &  GeV   & $x$         &  GeV              & GeV   \\ \hline \hline
\begin{tabular}{c} ESRF, PETRA-III \end{tabular}   &  6.0   &   0.22      &   1.09            &     1.43   \\\hline
  European XFEL       &  20 &    0.74     &     8.49          &     16.13    \\   \hline
HERA        &  26.5  &    0.98     &     13.1          &     23.4    \\   \hline
SLC, LEP    & 45.6   &     1.68    &     28.6          &    43.7 \\ \hline
ILC   & 250   &     9.23    &     226          &    249.6 \\
\hline
\end{tabular}
\caption{\label{tab1} Sensitivity of different accelerators' Compton facilities to the electron's refractive index.}
\end{table}
 Although HERA, SLC and LEP are not operational anymore, one can analyze
available data recorded on these accelerators 
for polarimetry studies. One can use the data to check if there is a deviation from predictions of GR affecting the Compton edge.
This is true for the HERA and SLC  but not for the LEP Compton polarimeter, which generated and registered many photons per machine pulse~\cite{LEP-polarimeter}. In this multi-photon regime, any shift of the Compton edge is convoluted with the laser-electron luminosity and cannot be disentangled and measured separately.

\begin{figure}[t]
\centering
\includegraphics[width=8cm]{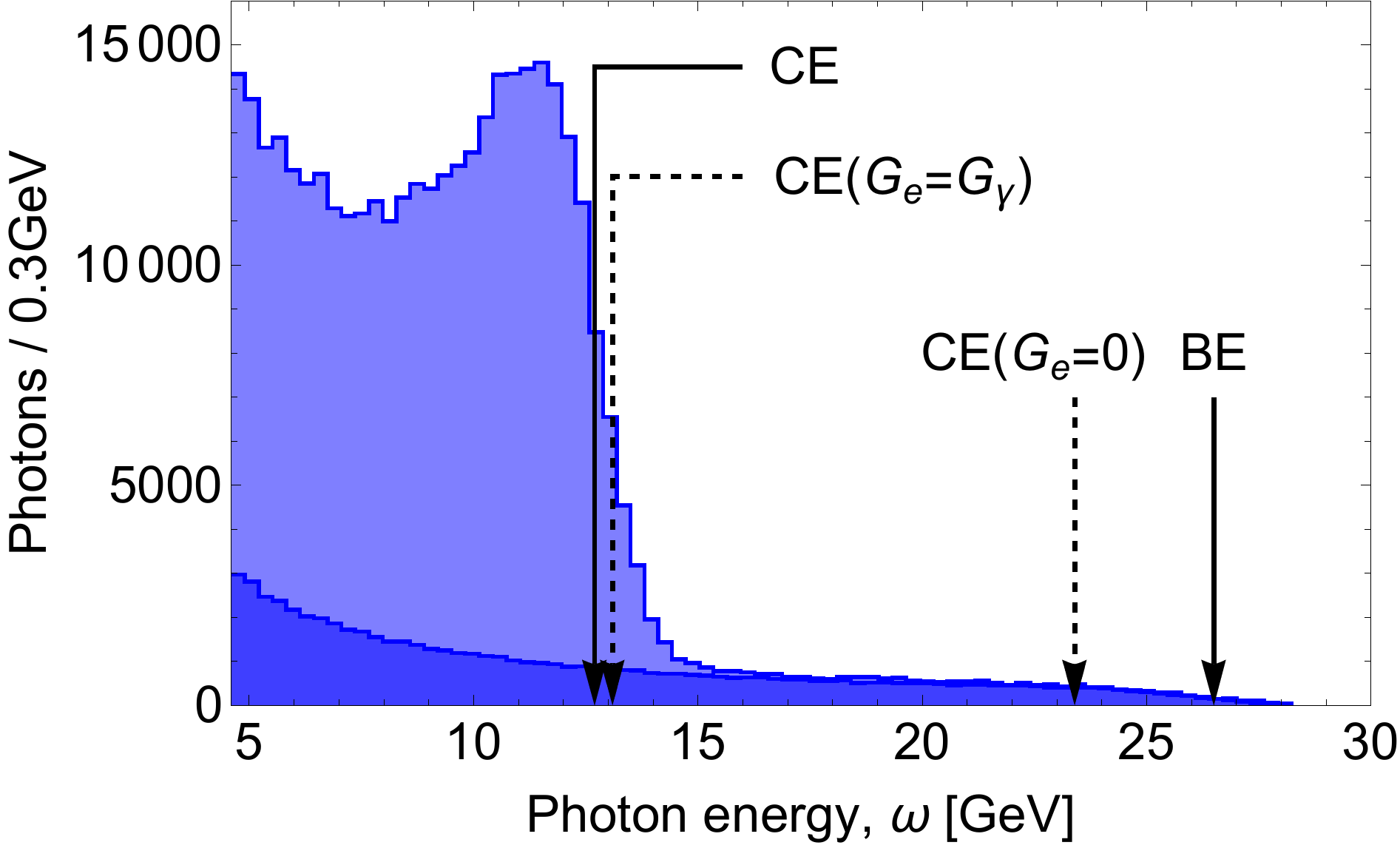}
\caption{\label{fig2}
HERA polarimeter Compton and Bremsstrahlung (darker area) spectra.
Vertical solid lines show measured positions of  the Compton (CE) and Bremsstrahlung (BE)
maximal energies. The dashed lines correspond to the predicted Compton edge for general relativity ($G_e=G_\gamma$) and intentionally changed ($G_e = 0$) case.
}
\end{figure}

\begin{figure*}[t!]
\includegraphics[width=8.5cm]{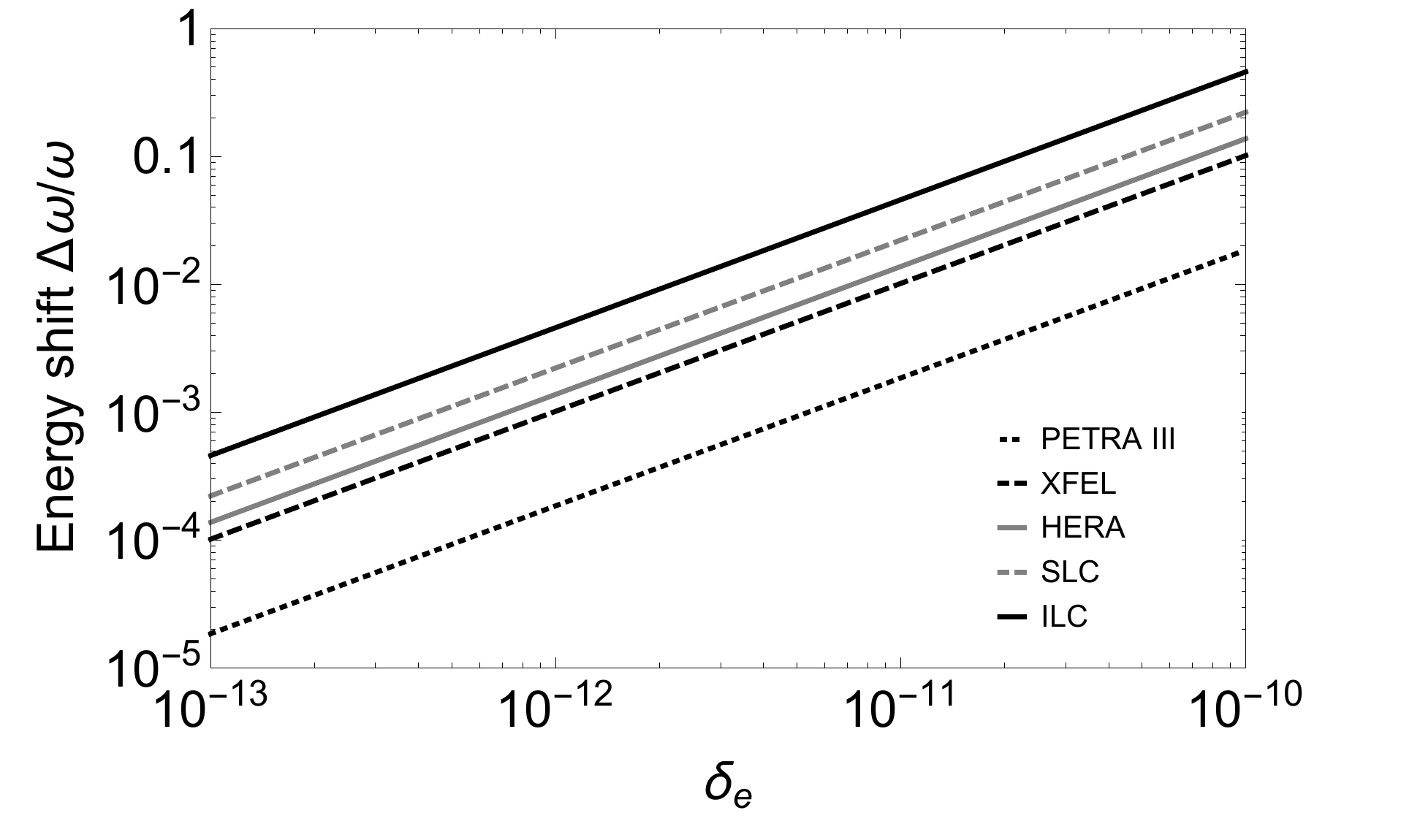}\hspace{0.5cm}
\includegraphics[width=8.4cm]{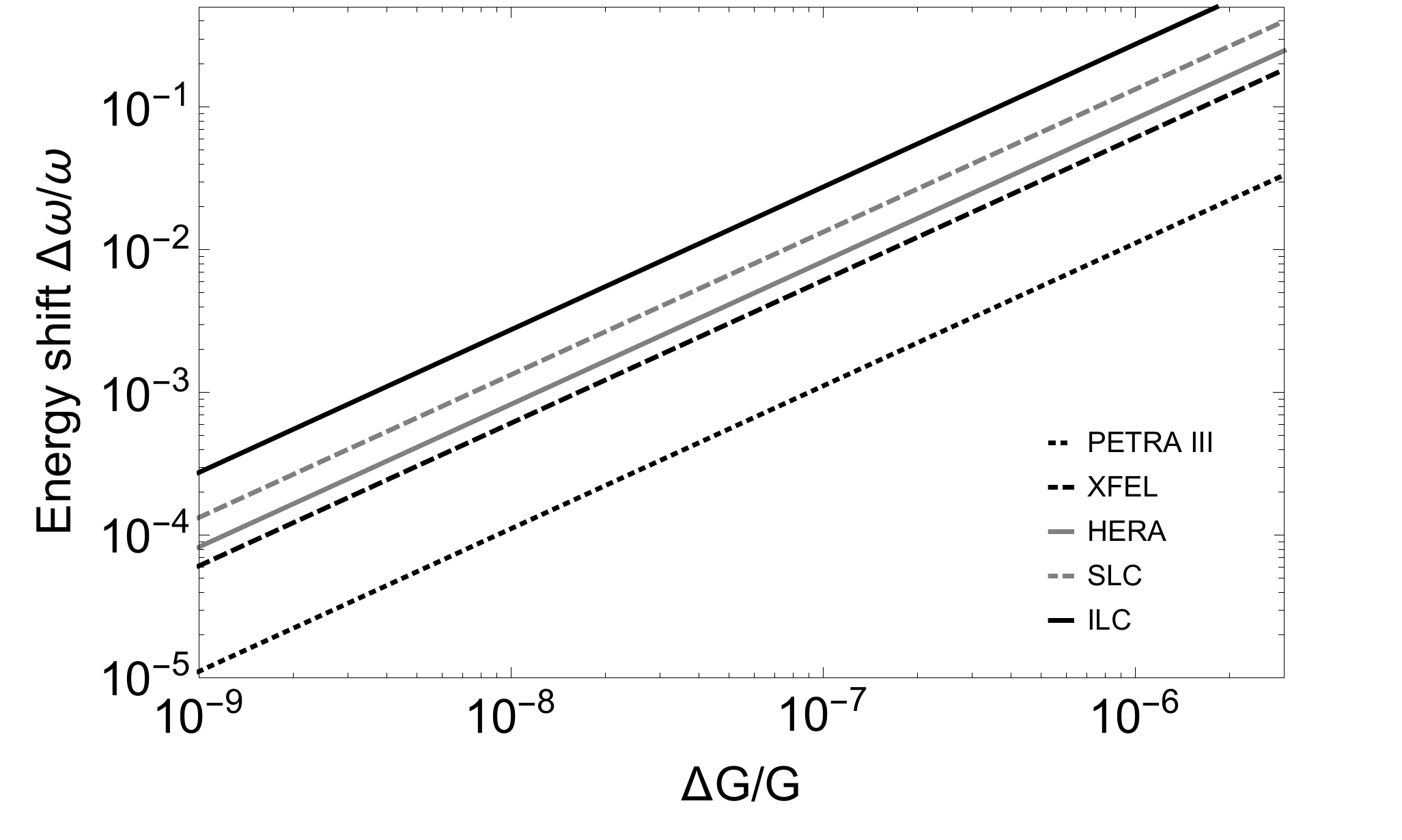}
\caption{\label{figne}
Sensitivity to the $\delta_e$ (left) and $\Delta G / G$ (right) deviations for the existing and future experiments. The $\Delta G / G$ was obtained from $\delta_e$ with the potential of the Local Supercluster (see the text).
}
\end{figure*}

Unlike the LEP, the SLC polarimeter operated in a multi-electron mode and analyzed
the energies of scattered electrons using a magnetic spectrometer~\cite{slac-pol}. The spectrometer
converted energies to positions, which then were detected by an array of Cherenkov counters.
Relationship between the position $S_x$ and the energy ${\cal E}' = {\cal E}-\omega$ of the scattered electron can be obtained from the spectrometer's magnetic field according to the expression
\begin{equation}
S_x =\frac{296.45~\mathrm{GeV}\cdot \mathrm{cm}}{{\cal E}'} - 9.61~\mathrm{cm}\,.
\label{eqhorz}
\end{equation}
The scaling factor is from Ref.~\cite{slac-pol}
and the offset, which depends on the electron beam position at the interaction point, corresponds to a calibration from Ref.~\cite{Gharibyan:2003fe}.
According to this  relation, the SLC polarimeter's Compton edge electrons with 17.4~GeV
energy will enter the detector at a position of 7.43~cm. This is what was measured
with $200 \upmu\mathrm{m}$ statistical accuracy by a kinematic endpoint scan and is presented
in Fig.~3-9 of Ref.~\cite{slac-pol}. This is a great accuracy, taken that the Compton edge electrons in the case $n_e=\beta$
(i.e. $G_e = 0$) and energy ${\cal E}'={\cal E}-\omega_{max} = 1.9$ GeV would have entered at a position of 146.4~cm.
Possible instrumental influence is limited to the initial electron beam position
shift, less than 1~cm (to be contained in the accelerator's magnetic
lattice~\cite{Chao:2013rba}) and
an estimated accuracy of the magnetic spectrometer, better than 2\%. These factors
add up to a maximum energy uncertainty or a possible offset of 1.4~GeV for the
measured value of 17.4~GeV, giving an upper bound on the refractive index deviation, $\bigl | \delta_e \bigr | < 2\cdot 10^{-11}$. We therefore conclude that the SLC polarimeter data does support the equivalence principle (and GR gravitational bending) with a good accuracy.

At the HERA transverse polarimeter, Compton photons are registered by a calorimeter
in a single particle counting mode. An example of the recorded Compton spectrum (adopted from Ref.~\cite{Barber:1992fc}) is shown in Fig.~\ref{fig2} together with a background Bremsstrahlung distribution.

In contrast to the Compton scattering, in the Bremsstrahlung process, the momentum
transfer is not fixed, and any small refractive effect is smeared out and becomes
negligible. Hence, we calibrate the energy scale according to the maximal Bremsstrahlung energy (see analysis in Ref.~\cite{Gharibyan:2003fe}) and show the experimental Compton edge energy in Fig.~\ref{fig2}, relative to
the Bremsstrahlung edge and the theoretical values. The figure shows a span of possible values of the Compton edge from $n_e=\beta$ to $n_e$ given by Eq.~(\ref{ne}). As it is visible from the plot, deviation of order $10^{-9}$ in the electron refractive index would create a mismatch of order of many GeV's in the position of the edge.
Comparing a measured maximal Compton energy of $12.7\pm0.1$~GeV from
Ref.~\cite{Gharibyan:2003fe}\footnote{Conclusions on the value of $n$ from that paper should be omitted, since $n_e$ was not taken into account} with the nominal 13.1~GeV opens a room for the speculations with a mismatch $\delta_e = 2\cdot 10^{-11}$.
However, there are instrumental errors \cite{Barber:1992fc} adding 1.3\% for the non-linear response of the calorimeter at given energies and at least 1\% for spatial non-uniformity of the calorimeter, possibly reducing the mismatch to $\delta_e = 4\cdot 10^{-12}$ or smaller. In addition, acceleration of the electron due to its electromagnetic interaction with the beam and vacuum chamber can be few percent comparing to its gravitational acceleration, which will give the right magnitude of effective $\Delta G$ and hence a shift in the Compton edge explaining the data (this analysis is out of scope of the article and will be presented elsewhere; for the estimates of the transverse forces see \cite{Chao:1993zn}). Therefore, we have to conclude that the HERA Compton experiment had a potential of either confirming or discarding the equivalence principle at high energies, if the systematic errors were studied more carefully.

In order to make a more concrete conclusion on whether a deviation from GR was indeed observed at HERA, we suggest to repeat the experiment at existing storage rings (e.g. PETRA-III) as well as at the future International Linear Collider (ILC) and European X-ray Free Electron Laser (XFEL), see parameters in Table~\ref{tab1}. Assuming $\Delta \omega / \omega_{max} \sim 10^{-3}$, one would be able to measure the possible mismatch with precision of $\delta_e \sim 5\cdot 10^{-12}$, $\delta_e \sim 10^{-12}$ and $\delta_e \sim 2\cdot 10^{-13}$ for PETRA-III, European XFEL and ILC, respectively, see Fig.~\ref{figne} (left).

In order to present the best sensitivity in terms of $\Delta G / G$ and use Eq.~(\ref{wtog}), we should identify the source of the largest gravitational potential at the surface of the Earth. Following Ref.~\cite{Hughes:1990ay} we replace the Earth's gravitational potential, $\Phi_\oplus = -G M_\oplus / R_\oplus = -7 \times 10^{-10}$ in (\ref{wtog}), by the gravitational potential of the Local Supercluster with $|\Phi_\mathrm{SC}| \simeq 3\times 10^{-5}$. Result is shown in Fig.~\ref{figne} (right). Taken the current bound on the graviton mass \cite{Agashe:2014kda}, $m_G < 6\times 10^{-32}$ eV, and, hence, the minimal range of the gravitational forces $\sim 100$~Mpc, one can improve our (conservative) estimates by taking into account gravitational potentials from larger or more distant mass distributions.

The use of absolute potentials is not a universally accepted practice and, hence, to make our results more robust, one should replace the absolute potentials by their gradients. Experimentally, this can be done by performing several Compton scattering measurements separated by time intervals sufficient to produce variations in the ambient gravitational potentials. One of examples of such variations is related to the changes in the distance between Earth and Sun due to the eccentricity of the Earth's orbit~\cite{Kalaydzhyan:2015ija}. Let us imagine that the measured Compton edge coincided with its nominal value (\ref{nominal}) in two experiments within the uncertainties $\Delta \omega_1$ and $\Delta \omega_2$, respectively. In this case, one can show that
\begin{align}
\left|\frac{\Delta G}{G}\right| < \frac{\Delta \omega_1 + \Delta \omega_2}{\omega_{max}}\cdot\frac{m^2 (1+x)^2}{4{\cal E}^2 |\Delta \Phi|}\,,
\end{align}
where $\Delta \Phi$ is the difference in the gravitational potentials around the accelerator for the two experiments. If $\Delta \omega_1 \simeq \Delta \omega_2$, then the strongest bounds can be placed by performing measurements at the moments when the Earth is at the perihelion (around January 3) and aphelion (around July 4) of its orbit. In this case, $|\Delta \Phi| = 2.43\times 10^{-10}$, see Ref.~\cite{Kalaydzhyan:2015ija}. At ILC, the polarimetry will be used as a real-time diagnostic tool and the Compton scattering will be performed with every bunch~\cite{Behnke:2013lya}, which makes it an ideal setup for our purposes. The day-to-day data on the position of the Compton edge has an additional advantage: finite $\Delta G$ would manifest itself in an annual variation of the data, which disentangles it from the time-independent systematic errors (possibly present in, e.g., HERA data).

{\it Conclusions.---}
In order to test the equivalence principle, 
we first described gravity effects in equivalent refractivity terms.
Next, we analyzed the high-energy laser Compton scattering,
which is extremely sensitive to any small refractivity due to its well-defined
initial and final energy states and fixed momentum transfers.
Finally, we explored available experimental records from the SLC and HERA
Compton polarimeters, finding that SLC confirms results of GR with a high accuracy,
while HERA suggests that relativistic electrons may behave slightly different from the GR predictions (within SLC bound).
However, due to the large (for this type of study) instrumental error for the HERA calorimeter and electromagnetic forces,
the results are not very conclusive, and we propose devoted experiments on existing and future facilities which would lead
to stronger tests of the theory of general relativity.

{\it Acknowledgements.---} This work was supported in part by the U.S. Department of Energy under Contract No. DE-FG-88ER40388. I would like to thank Vahagn Gharibyan, Martin Rocek and Jenny List for discussions.

\end{document}